\documentclass[
    reprint,aps,prl,twocolumn,nofootinbib,preprintnumbers
]{revtex4-1}

\usepackage[T1]{fontenc}
\usepackage[utf8]{inputenc}
\usepackage{lmodern, amsmath}
\usepackage{mathtools} 
\usepackage{cancel} 
\usepackage{MnSymbol} 
\usepackage{graphicx,adjustbox} 
\usepackage{mathrsfs} 
\usepackage{comment}

\usepackage{hyperref}
\hypersetup{
	colorlinks=true,
	linktoc=page,
	citecolor=americanrose,
	linkcolor=cadmiumgreen,
	urlcolor=blue} 
\def\appendixautorefname~#1\null{Appendix\,#1\null}
\def\sectionautorefname~#1\null{Section\,#1\null}
\def\subsectionautorefname~#1\null{Section\,#1\null}
\def\figureautorefname~#1\null{Figure\,#1\null}
\def\equationautorefname~#1\null{Equation\,(#1)\null}

\usepackage{xcolor}

\definecolor{americanrose}{rgb}{1.0, 0.01, 0.24}
\definecolor{cadmiumgreen}{rgb}{0.0, 0.42, 0.24}

\usepackage[colorinlistoftodos]{todonotes}
\setuptodonotes{color=cyan}
\presetkeys{todonotes}{inline}{}
\reversemarginpar 
\makeatletter\g@addto@macro\bfseries{\boldmath}\makeatother

\newcommand{\Res}{\mathrm{Res}}

\newcommand{\hdelta}{\hat{\delta}}

\newcommand{\oldornew}[1]{}

\begin{document}

\title{Spinning waveforms from KMOC at leading order}
\author{Stefano De Angelis}\email{stefano.de-angelis@ipht.fr}
\affiliation{Institut de Physique Théorique, CEA, CNRS, Université Paris-Saclay, F–91191 Gif-sur-Yvette cedex, France}
\author{Riccardo Gonzo}\email{rgonzo@exseed.ed.ac.uk}
\affiliation{Higgs Centre for Theoretical Physics, School of Physics and Astronomy, The University of Edinburgh, EH9 3FD, Scotland}
\author{Pavel P.~Novichkov}\email{pavel.novichkov@ipht.fr}
\affiliation{Institut de Physique Théorique, CEA, CNRS, Université Paris-Saclay, F–91191 Gif-sur-Yvette cedex, France}

\begin{abstract}
We provide the analytic waveform in time domain for the scattering of two Kerr black holes at leading order in the post-Minkowskian expansion and up to fourth order in both spins. The result is obtained by the generalization of the KMOC formalism to radiative observables, combined with the analytic continuation of the five-point scattering amplitude to complex kinematics. 
We use analyticity arguments to express the waveform directly in terms of the three-point coupling of the graviton to the spinning particles and the gravitational Compton amplitudes, completely bypassing the need to compute and integrate the five-point amplitude. 
In particular, this allows to easily include higher-order spin contributions for any spinning compact body. Finally, in the spinless case we find a new compact and gauge-invariant representation of the Kovacs--Thorne waveform.
\end{abstract}

\maketitle

\section{Introduction}

We are now living in the exciting era of gravitational-wave (GW) astronomy, with over 90 compact binary merger events detected to date by the LIGO-Virgo-KAGRA Collaboration \cite{LIGOScientific:2021djp}. Thanks to the improved sensitivity of the current and future GW detectors, many more events will be discovered in upcoming years. These ongoing searches for signals rely on analytical or numerical template banks both for detection and for parameter estimation. They call for a better theoretical understanding of gravitational waveforms.

Focusing on the inspiral phase, we can treat compact objects at large distances as point particles using an effective field theory approach to general relativity~\cite{Goldberger:2004jt,Porto:2016pyg}. Quantum field theory and amplitude-inspired techniques offer an analytic and efficient toolkit to perform classical calculations in the post-Minkowskian (PM) expansion \cite{Buonanno:2022pgc}, see the remarkable progress at 3PM and 4PM \cite{Bern:2019nnu,Bern:2021dqo,Bern:2021yeh,Bjerrum-Bohr:2021din,Brandhuber:2021eyq,Damgaard:2023ttc,Dlapa:2021npj,Dlapa:2022lmu,Kalin:2020fhe,FebresCordero:2022jts,Jakobsen:2022fcj,Jakobsen:2022zsx,Jakobsen:2023hig,Jakobsen:2023ndj}. Motivated by this, a framework to compute waveforms has been developed within the KMOC formalism \cite{Kosower:2018adc,Cristofoli:2021vyo} and the worldline approach \cite{Mogull:2020sak,Kalin:2020mvi}, albeit restricted so far to scattering configurations. Tree-level waveforms for spinless particles have been computed in  \cite{Jakobsen:2021smu,Mougiakakos:2021ckm,DiVecchia:2023frv} using the five-point amplitude \cite{Goldberger:2016iau,Luna:2017dtq,Bautista:2019evw}, making contact with the Kovacs and Thorne result obtained with traditional methods \cite{Kovacs:1977uw,Kovacs:1978eu} (see also the earlier result by Peters \cite{Peters:1970mx}). Recently, these calculations have been extended to one-loop order by combining KMOC with the heavy particle effective theory \cite{Brandhuber:2023hhy,Herderschee:2023fxh,Georgoudis:2023lgf,Elkhidir:2023dco}, whose results have been recently compared with post-Newtonian (PN) waveforms \cite{Bini:2023fiz} finding disagreement at higher order. The latter may be related to a classical part of the KMOC subtraction term, as stressed in \cite{Caron-Huot:2023vxl}.

Since astrophysical black holes are always spinning, the inclusion of spin effects is an important milestone in this program. A first step in this direction has been taken in \cite{Jakobsen:2021lvp} (see also \cite{Bautista:2021inx}), where effects quadratic in spin have been included in the tree-level PM waveform. To include higher-order spin effects, we need a full description of Kerr black holes in terms of spinning point particles. Such remarkable correspondence has been established first at the level of the three-point amplitude for the Kerr multipoles \cite{Vines:2017hyw,Arkani-Hamed:2017jhn,Guevara:2018wpp,Guevara:2019fsj,Chung:2018kqs,Aoude:2020onz}.
It is still under development for the four-point Compton amplitude \cite{Chiodaroli:2021eug,Aoude:2022trd,Aoude:2021oqj, Cangemi:2022bew,Saketh:2022wap,Alessio:2023kgf,Bautista:2023szu,Bern:2023ity,Haddad:2023ylx,Bjerrum-Bohr:2023iey}, which should match the conservative piece of the solution of the Teukolsky equation \cite{Bautista:2022wjf}. With such identification many conservative and radiative spinning amplitudes have been recently computed \cite{Vaidya:2014kza,Vines:2018gqi,Arkani-Hamed:2019ymq,Maybee:2019jus,Chung:2020rrz,Bern:2020buy,Kosmopoulos:2021zoq,Chen:2021kxt,Liu:2021zxr,Cristofoli:2021jas,Bastianelli:2021nbs,Aoude:2022thd,Aoude:2022thd,Alessio:2022kwv,Bern:2022kto,Riva:2022fru,Damgaard:2022jem,Menezes:2022tcs,Jakobsen:2022zsx,Jakobsen:2023hig,Jakobsen:2023ndj,Bianchi:2023lrg,Aoude:2023vdk,Bini:2023mdz,Heissenberg:2023uvo}.

In this letter, we develop a new method to perform the phase space integration for KMOC observables by making use of the analytic properties of amplitudes in the complex plane. Focusing on tree level, we compute for the first time \href{https://bitbucket.org/spinning-gravitational-observables/tree-level-waveform/}{the time-domain waveform for Kerr black holes up to fourth order in both spins}, using only the factorization channels of the five-point amplitude. The method presented here applies to generic field theories and beyond the classical limit. 

\paragraph{Conventions} We work in the signature $(+ - - -)$, using relativistic units $c=1$ and with $\kappa:=\sqrt{32 \pi G}$ where $G$ is the Newton constant. For convenience, we adopt the notation $\hat{\delta}^n(\cdot) \equiv (2 \pi)^n \delta^n(\cdot)$ and $\hat{\mathrm{d}}^n q \equiv \mathrm{d}^n q /(2 \pi)^n$. We take the graviton legs of the amplitudes to be outgoing.

\section{KMOC observables from the analytic properties of the S-matrix}

We follow the computation of the waveform presented in \cite{Cristofoli:2021vyo} and based on the previously developed observable-based formalism (KMOC) \cite{Kosower:2018adc} (see also \cite{Caron-Huot:2023vxl}). The strain at future null infinity is
\begin{align}
	\label{eq:strain}
	&h ( x)  =  \frac{\kappa}{8 \pi |\vec{x}\, | } \nonumber \\
 &\qquad \times \int_{0}^{\infty}\!
	\hat{\mathrm{d}} \omega
	\ \Big[   W(b; k^-)  e^{-i \omega u }  +   \left[ W(b; k^+) \right]^{\ast}  e^{i \omega u} \Big]\ ,
\end{align}
where $k^\mu = \omega n^\mu = \omega(1, \hat{x})$ with \(\hat{x} = \vec{x} / \lvert \vec{x} \rvert\), \(u = x^0 - \lvert \vec{x} \rvert\) is the retarded time, $W(b; k^\pm)$ is the helicity-dependent \textit{spectral waveform} of the emitted gravitational wave and $b$ is the impact parameter.  For the classical scattering of two massive particles, the spectral waveform can be computed from the S-matrix
\begin{equation}
	\label{eq:KMOCwaveform}
	i W(b,k^h)= \left\llangle \int\!\!\! \mathrm{d}\mu\, e^{i(q_1\cdot b_1 + q_2 \cdot b_2)}\ \mathcal{I}_{a_h} \right\rrangle\ ,
\end{equation}
where the double-angle brackets are understood as the classical limit of the expression inside:
\begin{equation}
	\hat{\delta}^{(D)} (q_1 + q_2 - k)\ \mathcal{I}_{a_h} = \langle p_1^\prime p_2^\prime  |  S^{\dagger}a_h(\vec{k}) S | p_1 p_2\rangle \ ,
\end{equation}
and the measure is defined by 
\begin{align}
	\label{eq:measureinnn}
	\mathrm{d}\mu = \left[\prod_{i=1,2} 
	\hat{\mathrm{d}}^D q_i\ \hdelta(-2  {p}_i\cdot q_i +q_i^2)
	\right]\,
	\hdelta^{D} (q_1 + q_2 - k) \ ,
\end{align}
with $q_i = p_i - p_i^\prime$.
At leading order in perturbation theory, this simplifies to
\begin{align}
	\label{eq:KMOCtree}
	\begin{split}
    W^{(0)}(b, k^h) = & \int\!\!\!\mathrm{d}\mu\, e^{i(q_1\cdot b_1 + q_2 \cdot b_2)}\, \mathcal{M}_{5,\rm cl}^{(0)}(q_1, q_2, k^h) \, ,	\end{split}
\end{align}
where $\mathcal{M}_{5,\rm cl}^{(0)}(q_1, q_2, k^h)$ denotes the tree-level scattering amplitude after taking the classical limit. For the purpose of the computation, we can set $b_1=b$ and $b_2=0$: indeed, the symmetric result will be recovered by performing a translation at the level of the time-domain waveform. This corresponds to
\begin{equation}
\label{eq:symmetrisation}
        b \to b_1 - b_2\ , \qquad u \to u - n\cdot b_2\ .
\end{equation}

A convenient way to perform the integrations in \eqref{eq:KMOCtree} in $D=4$ is discussed in \cite{Cristofoli:2021vyo}. After integrating out $q_2$ using the momentum-conserving delta function, one can parameterize the remaining integration variable $q_1$ as
\begin{align}
	\label{eq:qparam}
	q_1= z_1 v_1 + z_2 v_2 + z_v \tilde{v} + z_b \tilde{b}\, ,
\end{align}
where
\begin{align}
		v_1^\mu= \frac{p_1^\mu}{m_1}\, ,\quad  v_2^\mu= \frac{p_2^\mu}{m_2}\, , \quad\tilde{v}^\mu = \frac{v^\mu}{\sqrt{-v^2}}\, , \quad\tilde{b}^\mu = \frac{b^\mu}{\sqrt{-b^2}}\, ,
\end{align}
$p_i$ and $m_i$ are respectively the incoming momenta and the masses of the scattered objects,
and
\begin{align}
 v^\mu = \epsilon^{\mu \nu \rho \sigma} v_{1 \nu}  v_{2 \rho}  \tilde{b}_{\sigma}\, , \quad v^2 = - (\gamma^2-1)\,, \quad  \gamma = v_1 \cdot v_2 \,.
\end{align}
Choosing $b$ to be the asymptotic impact parameter, we also have $b\cdot v_1 = b\cdot v_2 =0$. The Jacobian is then $\mathrm{d}^4q = \sqrt{\gamma^2-1} \, \mathrm{d}^4 z$ and the measure becomes
\begin{align}
  \mathrm{d}\mu = \frac{\sqrt{\gamma^2-1} \, \mathrm{d}^4 z}{(4\pi)^2 m_1 m_2} \,
  \delta\left(z_1 + \gamma z_2\right) \delta\left(z_2 (\gamma^2-1) + w_2\right)\, ,
\end{align}
where we have defined
\begin{align}
	w_1 = v_1 \cdot k\,,  \qquad w_2 = v_2 \cdot k\,,
\end{align}
and we can safely neglect the shift \(q_i^2\) in the delta functions of \eqref{eq:measureinnn} at this order (alternatively, \textit{barred} variables $\bar{p}_i = p_i-q_i/2$ and $\bar{p}_i = \bar{m}_i \bar{v}_i$ can be used without any approximation).
Hence, \eqref{eq:KMOCtree} becomes
\begin{align}
	\label{eq:z_integrals}
		W^{(0)} &= \int\! \mathrm{d} z_v \mathrm{d} z_b \,  e^{ - i z_b \sqrt{-b^2} } \frac{\hat{\mathcal{M}}_{5,\rm cl}^{(0)}}{(4\pi)^2 m_1 m_2 \sqrt{\gamma^2-1}} \ ,
\end{align}
where
\begin{equation}
	\hat{\mathcal{M}}_{5,\rm cl}^{(0)} = \left.\mathcal{M}_{5,\rm cl}^{(0)}(q_1,q_2,k^h) \right|_{ z_1 =  \frac{\gamma}{\gamma^2-1} w_2\, , \ z_2 = -  \frac{1}{\gamma^2-1} w_2 }\, .
\end{equation}

\begin{figure}[t!]
	\includegraphics[width=\columnwidth]{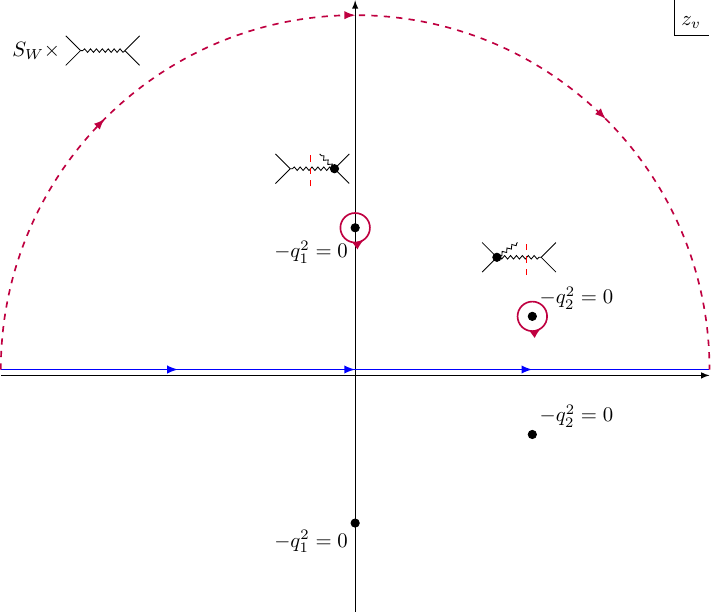}
	\caption{The deformation of the contour $\color{red}{\mathcal{C}^{(+)}}$ in the complex $z_v$ plane allows to evaluate the integral of the five-point tree-level amplitude directly in terms of the factorization channels.}
	\label{fig:analytic_continuation}
\end{figure}

We are going to evaluate the $z_v$ integral first, by deforming the contour of integration from the real axis to infinity through the upper half-plane (UHP) (or equivalently the LHP) as in \autoref{fig:analytic_continuation}. Having defined the integral
\begin{equation}
	I^{\lambda}_{\text{UHP}} = \int_{\color{red}{\mathcal{C}^{(+)}}} \! \mathrm{d} z_v \, \hat{\mathcal{M}}_{5,\rm cl}^{(0)} (q_1,q_2,k^{\lambda})\,, 
 \label{eq:UHP-zvintegral}
\end{equation}
we notice that it receives two different types of contributions from the residue theorem. The first is related to the simple poles of the tree-level five-point amplitude in the complex $z_v$-plane corresponding to $q_1^2=0$ and $q_2^2=0$ (the massive eikonal propagators take the form $w_i = v_i \cdot k$ and do not depend on $z_v$). These poles correspond to factorization channels and their residues are fixed as a product of lower-point tree-level amplitude summed over all the possible configurations of the internal states,
\begin{equation}
\begin{split}
\frac{I_{\mathcal{C}^{(+)}_{q_1}}^{\lambda}}{2 \pi i} &= \sum_{h} \mathcal{M}_{4, \mathrm{cl}}^{(0)}(p_2,k^{\lambda},-q_1^{h}) \mathcal{M}_{3, \mathrm{cl}}^{(0)}(p_1,q_1^{h}) \operatornamewithlimits{Res}_{z = \hat{z}_{1}} \frac{-1}{q_1^2} \,,\\
\frac{I_{\mathcal{C}^{(+)}_{q_2}}^{\lambda}}{2 \pi i} &= \sum_{h} \mathcal{M}_{4, \mathrm{cl}}^{(0)}(p_1,k^{\lambda},-q_2^{h}) \mathcal{M}_{3, \mathrm{cl}}^{(0)}(p_2,q_2^{h}) \operatornamewithlimits{Res}_{z = \hat{z}_{2}} \frac{-1}{q_2^2}\,,
\label{eq:Cq1q2_zv}
\end{split}
\end{equation}
where $\hat{z}_{1}$ and $\hat{z}_{2}$ are the solutions in the UHP of the pole constraints $-q_i^2 = (z_v - \hat{z}_i)(z_v - \hat{z}_i^*)=0$,  
\begin{equation}
\begin{split}
	\hat{z}_1 &= i \sqrt{z_b^2+w_{2,r}^2} \,,\\
	\hat{z}_2 &= -\tilde{v} \cdot k +i \sqrt{(z_b + \tilde{b}\cdot k)^2+w_{1,r}^2} \,,
 \label{eq:z1z2}
\end{split}
\end{equation}
which are written in terms of the rescaled variables
\begin{equation}
	w_{i,r} = \frac{w_i}{\sqrt{\gamma^2-1}}\ .
\end{equation}

An additional contribution is given by (half) the residue at infinity of the five-point amplitude. As proven in appendix \ref{sec:appendixA}, in the spinless case this is completely fixed by the Weinberg soft theorem \cite{Weinberg:1964ew} as the product of the leading soft factor and the four-point amplitude
\begin{align}
I_{\mathcal{C}^{(+)}_{\infty}}^{\lambda} &=  \int_{\mathcal{C}^{(+)}_{\infty}} \mathrm{d} z_v \, S_0(k^{\lambda}) \mathcal{M}_{4,\text{cl}}^{(0)} \,.
\label{eq:Cinf_cv}
\end{align}
This analysis also shows that the original $z_v$ integral in \eqref{eq:UHP-zvintegral} should be interpreted with a principal value prescription. On the other hand, the divergent pieces are proportional to non-negative powers of $z_b$ and they integrate to short-range interactions $\delta^{(n)}(\sqrt{-b^2})$, which are not relevant in the classical approximation. Summarizing, we have the following equality
\begin{equation}
	I_{\text{UHP}}^{\lambda}  = I^{\lambda}_{\mathcal{C}^{(+)}_{q_1}} + I^{\lambda}_{\mathcal{C}^{(+)}_{q_2}} + I^{\lambda}_{\mathcal{C}^{(+)}_{\infty}}\,, 
 \label{eq:IUHP}
\end{equation}
where the terms are defined in \eqref{eq:Cq1q2_zv} and \eqref{eq:Cinf_cv}.

An explicit calculation of \eqref{eq:IUHP} shows that the residues on the UHP \eqref{eq:Cq1q2_zv} introduce a linear dependence on the variable $\tilde{v} \cdot k$ in the denominator, see \eqref{eq:z1z2}. For the scalar case, such spurious singularity must cancel at the level of the integrated (parity-even) waveform. Nevertheless, such cancellation is not directly manifest when we combine the contributions from the two poles, given that the numerator may be a polynomial in $\tilde{v} \cdot k$.\footnote{For example, any term quadratic in $v^\mu$ in the numerator can be written in terms of the velocities and the impact parameter,
\begin{align}
    v^\mu v^\nu &= -(\gamma ^2-1) (\eta ^{\mu \nu}+b^{\mu } b^{\nu }) +2\gamma  v_2^{(\mu } v_1^{\nu )}-v_1^{\mu } v_1^{\nu}-v_2^{\mu} v_2^{\nu}\ . \nonumber
\end{align}
} 
To ensure that this happens, we can choose to split the $z_v$-integration into two equal pieces and deform the integration contour in the UHP and LHP. As a consequence, the contributions from the arcs at infinity cancel each other ($I^{\lambda}_{\mathcal{C}^{(+)}_{\infty}} = I^{\lambda}_{\mathcal{C}^{(-)}_{\infty}}$) and only the finite poles contribute:
\begin{align}
	\label{eq:zv_integral_toresidues}
	\hspace{-10pt}	\int_{-\infty}^{+\infty} \mathrm{d} z_v\ \hat{\mathcal{M}}_{5,\rm cl}^{(0)} &= \frac{1}{2} \left(I_{\text{UHP}}^{\lambda} - I_{\text{LHP}}^{\lambda}\right) \nonumber \\
  \hspace{-10pt} &= \frac{1}{2} \left(I^{\lambda}_{\mathcal{C}^{(+)}_{q_1}} + I^{\lambda}_{\mathcal{C}^{(+)}_{q_2}}  - I^{\lambda}_{\mathcal{C}^{(-)}_{q_1}} - I^{\lambda}_{\mathcal{C}^{(-)}_{q_2}} \right) \nonumber \\
 \hspace{-10pt} &= i \pi \sum_{i=1,2}\left(\underset{z_v=z_i}{\Res} \hat{\mathcal{M}}_{5,\rm cl}^{(0)} - \underset{z_v=z_i^*}{\Res} \hat{\mathcal{M}}_{5,\rm cl}^{(0)}\right) \,.
\end{align}

The computation of the waveform in the time domain becomes trivial. Indeed, it is convenient to perform the $\omega$ integration before taking the Fourier transform in $z_b$ to the impact-parameter space.%
\footnote{The latter integration is rather involved and evaluates to Bessel functions, while the time domain result simplifies to elementary functions. This was first observed at LO in \cite{Jakobsen:2021smu} and further understood for NLO and beyond in \cite{Brandhuber:2023hhy,Herderschee:2023fxh}.} Written in terms of dimensionless variables
\begin{equation}
	w_i \to \omega\, \bar{w}_i\ , \,\,\, \bar{w}_i = n \cdot v_i \,, \quad z_v \to \omega\, z_v \,, \quad z_b \to \omega\, z_b \,,
 \label{eq:scalingwz}
\end{equation} 
the amplitude scales as\footnote{Indeed, the amplitude has dimension $4-5=-1$, but in gravity we have to consider additional powers of $\kappa$ (which has dimension $-1$) and in the perturbative expansion we have $\mathcal{M}_{5,\rm cl}^{(L)} \sim \kappa^{3+2L}$.}\begin{equation}
\label{eq:omega_scaling_scalar}
	\mathcal{M}_{5,\rm cl}^{(L)} \to \omega^{-2+L} \mathcal{M}_{5,\rm cl}^{(L)}\,.
\end{equation}
Because the integration measure in equation \eqref{eq:z_integrals} using \eqref{eq:scalingwz} becomes $\mathrm{d}^2 z \to \omega^2 \mathrm{d}^2 z$, the $\omega$ integral \eqref{eq:strain} can only give (derivatives of) delta functions and principal value (PV) integrals:
\begin{equation}
\label{eq:delta_or_PV_timedomain}
	\delta^{(L)}(u+z_b \sqrt{-b^2})\ ,\qquad \mathrm{PV}^{(L)}\frac{1}{(u+z_b \sqrt{-b^2})}\ ,
\end{equation}
if we consider the real and imaginary parts of the amplitudes, respectively. At tree level only delta functions can appear because the tree-level amplitudes in physical kinematics are real,\footnote{At higher orders, instead, this method has to be modified for \textit{tail} effects which involve (exponentiated) logarithms of $\omega$ \cite{Blanchet:1993ec,Porto:2012as,Galley:2015kus}.} and therefore the strain becomes
\begin{equation}
\hspace{-7pt}h^{(0)}(x) = \frac{\kappa}{(4 \pi)^3 |\vec{x}| \sqrt{-b^2} } \frac{(I_{\text{UHP}}^{\lambda} - I_{\text{LHP}}^{\lambda})}{ 4 m_1 m_2 \sqrt{\gamma^2 - 1}} \Bigg|_{z_b = - u/\sqrt{-b^2}}\,,
\label{eq:tree-waveform}
\end{equation}
where we have performed the time-domain Fourier transform in \eqref{eq:strain} using the delta constraint $\delta(u+z_b \sqrt{-b^2})$.

\section{Tree-level waveform for Schwarzschild black holes}

We now compute the tree-level scattering waveform for spinless particles by combining \eqref{eq:tree-waveform}, \eqref{eq:z_integrals} and \eqref{eq:zv_integral_toresidues} with the following three-point and four-point amplitudes
\begin{align}
	\mathcal{M}_{3,\rm cl}^{(0)}(p, k) &= - \kappa \, m^2 (\varepsilon \cdot v)^2 \,, \nonumber \\
 \mathcal{M}_{4,\text{cl}}^{(0)}(p,k_1,k_2) &= \frac{\kappa^2 m^2}{q^2}\left(\frac{v \cdot F_1 \cdot F_2 \cdot v}{v \cdot k_1}\right)^2 \,,
\end{align}
where $F_i^{\mu \nu} = k_i^\mu \varepsilon_i^\nu -k^\nu \varepsilon_i^\mu$ are the (linearized) field strengths of the gravitons. In this case, the calculation is straightforward and the dependence on $v^\mu$ drops automatically after the manipulation in \eqref{eq:zv_integral_toresidues}. Finally, the scalar waveform at \textit{leading order} is obtained from equation \eqref{eq:tree-waveform} using the symmetrization \eqref{eq:symmetrisation}.

We notice that the residues in \eqref{eq:Cq1q2_zv} keep the physical poles at $w_i=0$ and introduce three additional singularities (in complex time) given by $1/\sqrt{1+T_i^2}$ and $1/S^2$ factors, where we used the variables introduced in the seminal papers of Kovacs and Thorne \cite{Kovacs:1977uw,Kovacs:1978eu}:
\begin{align}
	T_i &\coloneq \frac{\sqrt{\gamma^2-1} \, (u-b_i\cdot n)}{\sqrt{-(b_1-b_2)^2} \, \bar{w}_i}\ ,\\
	S^2 &\coloneq \frac{T_1^2-2 \gamma  T_1 T_2+T_2^2}{\gamma ^2-1} - 1\ .
\end{align}
The time variables $T_1$ and $T_2$ are the characteristic time scales of the acceleration of the particles, while $S$ encodes the relative spacetime difference of the two bodies in relation to the light-cone of the observer. The former singularities are already introduced by the $1/q_i^2$ poles, while the latter requires the factor $1/q_1^2 q_2^2$ and it is a signature of gravitational non-linearities. As already noticed in the original paper, there might be physical points for which $S^2=0$ but such singularity is spurious. Indeed, we have
\begin{equation}
	(\gamma^2-1) S^2 = - \prod_{\sigma=\pm}(\gamma +T_1 T_2 - \sigma \sqrt{\left(T_1^2+1\right) \left(T_2^2+1\right)})\ ,
\end{equation}
but it is worth noticing that we can always factorize the term $(\gamma +T_1 T_2 - \sqrt{\left(T_1^2+1\right) \left(T_2^2+1\right)})$ in the numerator, so that the spurious singularity vanishes manifestly.

Our result agrees with Kovacs and Thorne \cite{Kovacs:1978eu} (see also \cite{Jakobsen:2021smu}), after picking the frame 
\begin{equation}
	\begin{split}
		v_1^\mu&=(1,0,0,0)\ ,\\
		v_2^\mu&=(\gamma,\sqrt{\gamma^2-1},0,0)\ ,\\
		\tilde{b}^\mu&=(0,0,-1,0)\ ,\\
		n^\mu&=(1,\cos\theta,\sin\theta \cos\phi,\sin\theta \sin\phi)\ ,
  \label{eq:frame1}
	\end{split}
\end{equation}
and using the following polarization vectors:
\begin{equation}
	\begin{split}
		\varepsilon^\mu &= \partial_\theta n^\mu + i \frac{\partial_\phi n^\mu}{\sin\theta}\ ,\\
		\varepsilon_{+}^{\mu \nu} = \Re&\left(\varepsilon^\mu \varepsilon^\nu\right) \ , \qquad \varepsilon_{\times}^{\mu \nu} = \Im\left( \varepsilon^\mu \varepsilon^\nu\right)\ .
  \label{eq:frame2}
	\end{split}
\end{equation}
Our computational strategy automatically picks the BMS frame\footnote{To compare with Kovacs and Thorne and the recent work \cite{Jakobsen:2021smu}, we need to redefine the waveform imposing that it vanishes in the far (retarded) past $\tilde{h}(x) \coloneq h(x) - \lim_{u\rightarrow -\infty} h(x)$, which corresponds to a different BMS frame than the one chosen in \eqref{eq:BMS-frame}. Indeed, we have ignored terms proportional to $\delta (\omega)$ to the waveform in frequency space, which give a time-independent contribution after the Fourier transform and sets $\lim_{u\rightarrow -\infty} h(x) = 0$.
} in which
\begin{equation}
	\lim_{u\rightarrow +\infty} h^{(0)}(x) = - \lim_{u\rightarrow -\infty} h^{(0)}(x)\ ,
 \label{eq:BMS-frame}
\end{equation}
and the sum of these two terms matches the linear memory in equation~$(27)$ of \cite{Jakobsen:2021lvp}. We find then a new compact and manifestly gauge-invariant expression for the strain,
\begin{widetext}
\begin{equation}
\begin{split}
	h^{(0)} ( x)  =& \frac{G^2 m_1 m_2
	}{ 
	|\vec{x}\, | \sqrt{-b^2}} \frac{1}{\bar{w}_{1}^2 \bar{w}_{2}^2 \sqrt{1+T_{2}^2} \left(\gamma +\sqrt{\left(1+T_{1}^2\right) \left(1+T_{2}^2\right)}+T_{1} T_{2}\right)}\\
	&\times \Bigg( \frac{3 \bar{w}_{1} + 2\gamma \left(2 T_1 T_2 \bar{w}_{1}-T_2^2 \bar{w}_{2}+\bar{w}_{2}\right)-\left(2 \gamma ^2-1\right) \bar{w}_{1}}{\gamma ^2-1}f_{1,2}^2 \\
	&-\frac{4\gamma T_2 \bar{w}_{2} f_{1}+2\left(2 \gamma ^2-1\right) \left[T_1 \left(1+T_2^2\right) \bar{w}_{2} f_{1}+T_2 (T_1 T_2 \bar{w}_{1}+\bar{w}_{2}) f_{2}\right]}{ \sqrt{\gamma ^2-1}} f_{1,2} \\
	&+4 \left(1+T_2^2\right) \bar{w}_{2}f_{1} f_{2}-4 \gamma  \left(1+T_2^2\right) \bar{w}_{2} \left(f_{1}^2+f_{2}^2\right)+2 \left(2 \gamma ^2-1\right) \left(1+2 T_2^2\right) \bar{w}_{2} f_{1} f_{2} \Bigg) + \left(1\leftrightarrow 2\right)\ ,
\end{split}
\end{equation}
\end{widetext}
where we have defined
\begin{align}
	f_{1,2} &= v_1 \cdot \varepsilon_k\, v_2 \cdot n - v_1 \cdot n\, v_2 \cdot \varepsilon_k\ ,\\
	f_{i} &= \tilde{b} \cdot \varepsilon_k\, v_i \cdot n - \tilde{b} \cdot n\, v_i \cdot \varepsilon_k\ .
\end{align}
It is worth emphasizing that these three structures are not independent of each other, but they are related (linearly) by the Bianchi identity
\begin{equation}
	f_{1,2} (T_{1} \bar{w}_{1}-T_{2} \bar{w}_{2})+\sqrt{\gamma ^2-1} (f_{2} \bar{w}_{1}-f_{1} \bar{w}_{2}) = 0\ ,
\end{equation}
and (quadratically) by the four-dimensional identity
\begin{equation}
	\hspace{-9pt}(f_{2} \bar{w}_{1}-f_{1} \bar{w}_{2})^2 =\frac{\left(f_{1}^2-2 \gamma  f_{1} f_{2}+f_{2}^2\right) (T_{1} \bar{w}_{1}-T_{2} \bar{w}_{2})^2}{\gamma ^2-1}\,.
\end{equation}

\section{Tree-level waveform for Kerr black holes}
\label{sec:Kerr}

In this section, we discuss the calculation of the scattering waveform for two Kerr black holes, which are identified with spinning point particles. The essential ingredients in the computation are the three-point and the four-point (Compton) amplitudes; although they are typically presented in the literature as helicity amplitudes \cite{Vines:2017hyw,Guevara:2018wpp,Guevara:2019fsj,Bautista:2019tdr}, we find more convenient for our purposes to use an equivalent gauge-invariant representation in terms of polarization vectors.

The gauge-invariant three-point amplitude describing the coupling of a Kerr black hole to gravity takes an exponential form \cite{Chung:2018kqs,Arkani-Hamed:2019ymq,Guevara:2018wpp}
\begin{equation}
    \label{eq:Kerr_coupling}
    \mathcal{M}^{(0)}_{3,\rm cl} = - \kappa \, m^2 \, (\varepsilon \cdot v)^2 \, \exp\left( \frac{i \, \varepsilon \cdot S \cdot k}{\varepsilon \cdot v} \right)\ ,
\end{equation}
where $S$ is the (mass-normalized) spin tensor of the Kerr black hole, related to the Pauli--Lubanski pseudo-vector $a^\mu$ by
\begin{equation}
    S^{\mu \nu} = \epsilon^{\mu \nu \rho \sigma} v_{\rho} a_{\sigma}\, , \qquad a^{\mu} = \frac{1}{2} \epsilon^{\mu \nu \rho \sigma} v_{\nu} S_{\rho \sigma}\, .
\end{equation}
The exponential features a spurious pole in the polarization vector~$\varepsilon$ starting at $\mathcal{O}(S^{3})$, which makes this expression unsuitable for computing higher-point amplitudes from unitarity \cite{Chung:2018kqs}.
Interestingly, the spurious pole can be removed by exploiting the four-dimensional identity
\begin{equation}
    \left(\frac{\varepsilon \cdot S \cdot k}{\varepsilon \cdot v}\right)^2 = k \cdot S \cdot S \cdot k = - (k \cdot a)^2\, .
\end{equation}
Expanding the exponential in \eqref{eq:Kerr_coupling} and applying this identity, we find an alternating structure
(which resums to sine and cosine~\cite{Bern:2020buy,Bjerrum-Bohr:2023jau})
free from the spurious pole in $\varepsilon$ at any order in the spin expansion (see Appendix~A of~\cite{Bautista:2019tdr}):
\begin{align}
    \mathcal{M}^{(0)}_{3,\rm cl} = & - \kappa \, m^2 \left[ (\varepsilon \cdot v)^2 + i \, (\varepsilon \cdot v) \, (\varepsilon \cdot S \cdot k )\right. \nonumber \\
    &  -\frac{1}{2} (\varepsilon \cdot v)^2 (k \cdot S \cdot S \cdot k) \nonumber \\
    & - \frac{i}{3!} (\varepsilon \cdot v) \, (\varepsilon \cdot S \cdot k) \, (k \cdot S \cdot S \cdot k) \nonumber \\
    & \left. + \frac{1}{4!} (\varepsilon \cdot v)^2 (k \cdot S \cdot S \cdot k)^2 + \dots \right]\ .
\end{align}

The gauge-invariant Compton amplitude is~\cite{Bautista:2019tdr,Bautista:2022wjf}
\begin{equation}
  \label{eq:Kerr_S2}
  \hspace{-5pt} \mathcal{M}^{(0)}_{4,\rm cl} = \frac{\kappa^2 m^2 \omega_0^2}{8 (k_1 \cdot k_2) (k_1 \cdot v)^2} \left( 1 + \frac{\omega_1}{\omega_0} + \frac{\omega_2}{\omega_0} \right) + \mathcal{O}(S^3) \, ,
\end{equation}
where $\omega_i$ are suitably contracted spin multipole coefficients
\begin{align}
   \hspace{-3pt} \omega_0 &= -2\, v \cdot F_1 \cdot F_2 \cdot v\, , \nonumber \\
    \hspace{-3pt} i\, \omega_1 &= k_1 \cdot F_2 \cdot v \, S \cdot F_1 - \frac{(k_1 - k_2) \cdot v}{2} S \cdot F_1 \cdot F_2 + \left( 1 \leftrightarrow 2\right)\, , \nonumber \\
   \hspace{-3pt} \omega_2 &= (k_1\cdot k_2) \bigg[ S \cdot S \left( \frac{F_1 \cdot F_2}{2} - v \cdot F_1 \cdot F_2 \cdot v \right) - 2\, S \cdot S \cdot F_1 \cdot F_2 \bigg] \nonumber \\
    & \qquad \qquad  - \frac{\omega_0}{2} (k_1 + k_2) \cdot S \cdot S \cdot (k_1 + k_2)\ ,
  \end{align}
where we define $A \cdot B = \eta_{\mu \nu} \eta_{\alpha \beta} A^{\alpha \mu} B^{\nu \beta}$ and $A \cdot B \cdot C \cdot D =  \eta_{\mu_1 \nu_1} \eta_{\mu_2 \nu_2} \eta_{\mu_3 \nu_3} \eta_{\alpha \beta} A^{\alpha \mu_1} B^{\nu_1 \mu_2} C^{\nu_2 \mu_3} D^{\nu_3 \beta} $ for any tensors $A,B,C,D$.
The gauge-invariant amplitude in \eqref{eq:Kerr_S2} can be equivalently written as an exponential for any graviton polarization (as recently noticed \cite{Bjerrum-Bohr:2023iey})
\begin{equation}
  \label{eq:Kerr_exp}
  \mathcal{M}^{(0)}_{4,\rm cl} = \frac{\kappa^2 m^2 \omega_0^2}{8 (k_1 \cdot k_2) (k_1 \cdot v)^2} \exp \left( \frac{\omega_1}{\omega_0} \right) + \mathcal{O}(S^5) \, ,
\end{equation}
and in four dimensions one has
\begin{equation}
  \label{eq:omega_id}
  \frac{\omega_2}{\omega_0} = \frac{1}{2} \left(\frac{\omega_1}{\omega_0}\right)^2\ .
\end{equation}

Similarly to the three-point case, the exponential~\eqref{eq:Kerr_exp} features a spurious pole in $\omega_0$ starting at $\mathcal{O}(S^3)$, which we remove by means of the identity~\eqref{eq:omega_id}, thus arriving at a manifestly gauge-invariant expression for the Compton amplitude valid up to $\mathcal{O}(S^4)$ and free of spurious poles:
\begin{equation}
  \label{eq:Kerr_compton}
   \hspace{-8pt} \mathcal{M}^{(0)}_{4,\rm cl} = \kappa^2 m^2 \frac{\omega_0^2 + \omega_0 \omega_1 + \omega_0 \omega_2 + \frac{\omega_1 \, \omega_2}{3} + \frac{\omega_2^2}{6}}{8 (k_1 \cdot k_2) (k_1 \cdot v)^2} + \mathcal{O}(S^5) \,.
\end{equation}
This amplitude matches the solution of the Teukolsky equation \cite{Bautista:2022wjf} and gives the correct factorization channel reproducing \eqref{eq:Kerr_coupling}.

Equipped with the amplitudes~\eqref{eq:Kerr_coupling} and \eqref{eq:Kerr_compton}, we repeat the same steps as in the scalar scattering case to compute the waveform up to the fourth order in spin for both black holes, i.e. $\mathcal{O}(S_1^4,S_2^4)$. Some additional considerations have to be made for the interactions generated by the spin multipole expansion. First, after applying the rescaling \eqref{eq:scalingwz}, the degree of homogeneity in $\omega$ is not the one predicted in \eqref{eq:omega_scaling_scalar}. The mass-dimension of $S_i^{\mu \nu}$ is $-1$ and if we write the five-point amplitude in the spin-multipole expansion \begin{align}
  \mathcal{M}^{(L)}_{5,{\rm cl}} &= \sum_{s_1,s_2=1}^{\infty} \left[ S_1^{\mu_1 \nu_1} \dots S_1^{\mu_{s_1} \nu_{s_1}} \right] \left[S_2^{\rho_1 \sigma_1} \dots S_2^{\rho_{s_2} \sigma_{s_2}} \right] \nonumber \\
 & \qquad \qquad \qquad \times \mathcal{M}^{(L),(s_1,s_2)}_{\mu_1 \nu_1, \dots \mu_{s_1} \nu_{s_1}, \rho_1 \sigma_1, \dots \rho_{s_2} \sigma_{s_2}}\ ,
 \end{align}
the scaling will be
\begin{equation}
  \mathcal{M}^{(L),(s_1,s_2)} \rightarrow \omega^{-2+L+s_1+s_2} \mathcal{M}^{(L),(s_1,s_2)} \, .
\end{equation}
Consequently, the recipe to compute the tree-level spin-multipole expansion of the waveform
\begin{equation}
	h^{(0)}(x) = \sum_{s_1,s_2=0}^\infty h^{(s_1,s_2)}(x)\, ,
\end{equation}
needs to be modified because the $z_v$ integral gives contributions of the type $\sim \delta^{(s_1+s_2)}(\sqrt{-b^2} z_b + u)$. In Appendix~\ref{sec:appendixB}, we discuss how the scaling of the spinning tree-level amplitudes at large $z_v$ affects the evaluation of the waveform, giving in particular additional terms corresponding to negligible short-range interactions. This shows that the spinning tree-level waveform can be directly computed with the following prescription
\begin{widetext}
\begin{equation}
    \hspace{-2pt}h^{(s_1,s_2)}(x) = \frac{\kappa}{(4 \pi)^3 \, |\vec{x}| \, (\sqrt{-b^2})^{s_1+s_2+1}} \frac{(-i)^{s_1+s_2}}{4 m_1 m_2 \sqrt{\gamma^2 - 1}} \frac{\partial^{s_1+s_2}}{\partial^{s_1+s_2} z_b} (I_{\text{UHP}}^{(s_1,s_2)} - I_{\text{LHP}}^{(s_1,s_2)}) \Bigg|_{z_b = - u/\sqrt{-b^2}}\,,
\label{eq:tree-waveform-spin}
\end{equation}
\end{widetext}
where the residues are now computed isolating the $s_1^{\rm th}$ and $s_2^{\rm th}$ multipoles in the factorized amplitudes. 

The final result can be written as 
\begin{equation}
	h^{(0)}(x) = \sum_{s_1,s_2=0}^{4} \frac{G^2 m_1 m_2}{|\vec{x}| (\sqrt{-b^2})^{s_1+s_2+1} } \mathfrak{h}_{s_1,s_2}(x) + \mathcal{O}(S_1^5,S_2^5)\, ,
\end{equation}
where the $\mathfrak{h}_{s_1,s_2}$'s are provided in \href{https://bitbucket.org/spinning-gravitational-observables/tree-level-waveform/}{ancillary files}. For $s_1+s_2 \leq 2$, our results match the ones presented in~\cite{Jakobsen:2021lvp}.

\section{Conclusion}

In this letter, we combined the KMOC formalism with the analytic properties of the S-matrix to develop an efficient framework for the calculation of the time-domain gravitational waveform for spinless and spinning bodies.

Observables like the waveform involve a phase-space integration over the classical amplitude, which as we showed can be easily evaluated by complex analysis tools once its singularity structure is understood. Focusing on the leading order, we computed the time-domain waveform directly from the factorization channels of the five-point amplitude, bypassing the complexity in its direct calculation. This is particularly convenient in the spinning case, because we only need the three-point and the four-point (Compton) amplitudes to determine the waveform solely algebraically. 

Using our method, we first provided a new compact gauge-invariant expression of the leading-order waveform for spinless particles, discussing in detail its singularity structure. Our result agrees with the traditional Kovacs--Thorne result \cite{Kovacs:1978eu} and a more recent worldline calculation \cite{Jakobsen:2021lvp}. We then considered the spinning case, where we use a new gauge-invariant representation of the Compton amplitude which agrees with the solution of the Teukolsky equation \cite{Bautista:2022wjf} to compute analytically the leading-order waveform relevant for scattering of Kerr black holes. At quadratic order in spin this agrees with ref.\cite{Jakobsen:2021lvp}. We provided an analytic expression valid up to fourth order in spin which can be directly extended to all spin orders once the full Kerr Compton amplitude is understood.

We leave a number of open questions to future investigations. The first is to understand to which extent classical observables depend on the analytic structure of amplitudes (i.e., poles and branch cuts), which as shown in this work can help to bypass traditional techniques. A second pressing problem is to understand the analytic continuation for the waveform discussed here, building on the dictionary between scattering and bound observables \cite{Turner:1977,Damour1985,Blanchet:1989,Kalin:2019rwq,Kalin:2019inp,Bini:2020hmy,Cho:2021arx,Saketh:2021sri,Bautista:2021inx,Adamo:2022ooq,Gonzo:2023goe}. An additional future direction is moving away from the restriction of having Kerr black holes and consider more general compact spinning objects (like neutron stars), allowing generic multipoles for the three-point coupling \cite{Arkani-Hamed:2017jhn}. In such case, the four-point amplitudes can be bootstrapped imposing locality and unitarity, up to contact interactions which can be taken into account properly.
Finally, it would be interesting to compare directly post-Minkowskian scattering waveforms both with analytic post-Newtonian spinning waveforms \cite{Junker:1992,Kidder:1995zr,Majar:2010em,Blanchet:2013haa,DeVittori:2014psa,Cho:2018upo,Bautista:2021inx} and effective-one-body waveforms \cite{Damour:2014afa,Nagar:2020xsk}, with the idea that in the future we might be able to detect black-hole hyperbolic encounters \cite{Capozziello:2008ra,Mukherjee:2020hnm,Kocsis:2006hq}.

\paragraph{Acknowledgments} We thank Fabian Bautista, Gustav Jakobsen, David Kosower,  Gustav Mogull, Jan Plefka and Massimiliano Riva for useful discussions and Rafael Aoude, Kays Haddad and Andreas Helset for communication on the related work \cite{Aoude:2023dui}. SDA and PPN's research is supported by the European Research Council, under grant ERC–AdG–88541. RG would like to thank FAPESP grant  2021/14335-0 where part of this work was done during the month of August 2023.

\paragraph{Note added} Shortly after this work was posted on the arXiv, the preprints \cite{Brandhuber:2023hhl,Aoude:2023dui} appeared. In reference \cite{Brandhuber:2023hhl}, the authors apply the integration method presented in this paper to the study of spinless-spinning scattering using the resummed-in-spin Compton amplitude presented in \cite{Bjerrum-Bohr:2023iey}, which matches the result from black-hole perturbation theory up to $\mathcal{O}(a^4)$. In reference \cite{Aoude:2023dui}, the authors computed the waveform with the generic parametrisation of the spin multipoles developed in \cite{Aoude:2020onz,Aoude:2022trd,Aoude:2022thd,Aoude:2023vdk}. Both approaches successfully reproduced our results.

\newpage

\appendix

\section{Residue at infinity and soft theorems}
\label{sec:appendixA}

In this appendix, we show that the contribution at infinity to the $z_v$ integral in \eqref{eq:UHP-zvintegral} is completely determined by the leading soft theorem in the classical limit. Indeed, when we consider the large-$z_v$ limit, we have $q_1 \sim -q_2 \sim z_v \tilde{v} \gg k$. Because of the heavy-mass (or classical) expansion we also have $p_i \gg q_i$. Then, combining the two conditions, this corresponds exactly to taking $k^\mu$ soft (i.e. $k \ll p_i, q_i$). In the soft limit $k^{\mu} \to 0$, the five-point amplitude becomes \cite{Weinberg:1964ew,Bautista:2019evw,Brandhuber:2023hhy}
\begin{equation}
\begin{split}
	\hspace{-7pt}&\qquad \, \mathcal{M}_{5,\rm cl}^{(0)}(q,k^{\lambda}) = S_0(k^{\lambda}) \mathcal{M}_{4,\rm cl}^{(0)}(q) + \mathcal{O}(\omega^0) \,,  \\
  \hspace{-7pt}&S_0(k^{\lambda}) = \frac{\kappa}{2}\frac{p_1\cdot F\cdot p_2}{(p_1\cdot k)(p_2\cdot k)} \left( \frac{p_1\cdot F\cdot q}{p_1\cdot k}  + \frac{p_2\cdot F \cdot q}{p_2\cdot k}\right) \,,
 \label{eq:soft-limit}
\end{split}
\end{equation}
where $q^\mu = q_1^\mu = - q_2^\mu$. The large-$z_v$ limit corresponds to
\begin{equation}
	\begin{aligned}
		p_1\cdot k\sim z_v^0\,, \quad p_i \cdot F\cdot q \sim z_v^1\ ,
	\end{aligned}
\end{equation}
which then gives
\begin{equation}
\begin{aligned}
	\hspace{-5pt}\mathcal{M}_{5,\rm cl}^{(0)} &= \frac{\kappa}{2}\frac{p_1\cdot F\cdot p_2}{(p_1\cdot k)(p_2\cdot k)} \\
	\hspace{-5pt}&\times \left( \frac{p_1\cdot F\cdot \tilde{v}}{p_1\cdot k}  + \frac{p_2\cdot F \cdot \tilde{v}}{p_2\cdot k}\right) z_v \mathcal{M}^{(0)}_{4,\rm cl} + \mathcal{O}(z_v^{-2})\,,
\end{aligned}
\end{equation}
where we have noticed that
\begin{equation}
    \mathcal{M}^{(0)}_{4,\rm cl}\sim |q|^{-2} \sim 1/z_v^2\ .
\end{equation}
Therefore, at tree level the five-point amplitude goes as $z_v^{-1}$ for large $z_v$ and the principal value prescription makes the integral finite.

\section{The regulator prescription for spinning waveforms}
\label{sec:appendixB}

We explain here the refined principal value prescription we use to compute the spinning tree-level waveform. The mass scaling of the Pauli--Lubanski pseudo-vector tells us that the large-$z_v$ behavior of the four-point amplitude is modified in the spin-multipole expansion
\begin{equation}
    \mathcal{M}^{(0),(s_1,s_2)}_{4,\rm cl}\sim |q|^{-2+s_1+s_2} \sim z_v^{-2+s_1+s_2}\ .
\end{equation}
This means that the $z_v$ integral of $\mathcal{M}_{5, \mathrm{cl}}^{(0),(s_1,s_2)} \sim z_v^{-1+s_1+s_2}$ diverges when $s_1+s_2>0$ even with the principal value prescription proposed in the scalar case.%
\footnote{It may be worth noting that such divergence emerges from taking the classical limit before integration.} On the other hand, there is a regulating procedure motivating \eqref{eq:tree-waveform-spin} using distributional identities. In particular, we can swap the integration order as follows: 
\begin{align}
   \hspace{-3pt} h^{(0)}(x) & \sim \int\!\! \mathrm{d} z_v \!\int\!\! \mathrm{d} z_b \!\int\!\! \mathrm{d}\omega\, e^{- i \omega (z_b \sqrt{-b^2}+u)} \omega^{-2+s_1+s_2} \mathcal{M}_{5,\rm cl}^{(0)} \nonumber\\
   \hspace{-3pt} & \sim \int\!\! \mathrm{d} z_v \!\int\!\! \mathrm{d} z_b \ \delta^{(s_1+s_2)}(z_b \sqrt{-b^2}+u)\, \mathcal{M}_{5,\rm cl}^{(0)} \nonumber \\
   \hspace{-3pt} & \sim \int\!\! \mathrm{d} z_v \ \partial^{(s_1+s_2)}_{z_b} \mathcal{M}_{5,\rm cl}^{(0)} \Big|_{z_b = - u/\sqrt{-b^2}} \nonumber \\
   \hspace{-3pt} & \sim b_{\mu_1} \cdots b_{\mu_{s_1+s_2}} \int\!\!\mathrm{d} z_v \ \partial_{q_1}^{\mu_1 \dots \mu_{s_1+s_2}} \mathcal{M}_{5,\rm cl}^{(0)} \Big|_{z_b = - u/\sqrt{-b^2}}\ ,
   \label{eq:distributional}
\end{align}
where all the variables are rescaled according to \eqref{eq:scalingwz}. At this point, we notice that the derivative w.r.t. $q_1^\mu$ scales as $z_v^{-1}$ in the large-$z_v$ limit, therefore compensating the behavior of the amplitude, so that the integrand scales as $z_v^{-1}$ like in the scalar case. Therefore, after using the distributional identities in \eqref{eq:distributional}, we can adopt the principal value prescription for the $z_v$ integral. The result is then indeed finite and well-defined.

\bibliographystyle{apsrev4-1_title}
\bibliography{binary}

\end{document}